\documentclass[a4paper,11pt]{article}
\usepackage{jinstpub} 
\usepackage{lineno}
\usepackage{siunitx}
\usepackage[subrefformat=parens]{subcaption}
\usepackage{xcolor}
\usepackage{ulem}

\newcommand{\erase}[1]{\if0{#1}\fi}
\newcommand{\add}[1]{\textcolor{black}{#1}}

\title{\boldmath Development and performance evaluation of a thin  GAGG:Ce scintillator plate for high resolution synchrotron radiation X-ray imaging}

\author[a,b]{Masao Yoshino}
\author[c]{Seiichi Yamamoto}
\author[d]{Kohei Nakanishi}
\author[d]{Katsunori Yogo}
\author[a,b]{Kei Kamada}
\author[c]{Nanase Koshikawa}
\author[c]{Jun Kataoka}
\author[a,b,e]{Akira Yoshikawa}

\affiliation[a]{New Industry Creation Hatchery Center, Tohoku University, 6-6-10, Aoba, Aramaki, Aoba-ku, Sendai, Miyagi, Japan}
\affiliation[b]{C\&A Corporation, 1-16-23 Ichibancho, Aoba-ku, Sendai, Miyagi, Japan}
\affiliation[c]{Fuculty of Science and Engineering, Waseda University, 3-4-1, Okubo, Shinjuku-ku, Tokyo, Japan}
\affiliation[d]{Nagoya University Graduate School of Medicine, 1-1-20 Daiko-minami, Higashi-ku, Nagoya, Aichi, Japan}
\affiliation[e]{Institute for Materials Research, Tohoku University, 2-1-1, Katahira, Aoba-ku, Sendai, Miyagi, Japan}
\emailAdd{masao.yoshino.a5@tohoku.ac.jp}

\abstract{
Scintillator-based X-ray imaging detectors are pivotal in numerous scientific and practical domains, including medical imaging, food and device inspection, and security monitoring. Recent advancements have spurred interest in 4D X-ray imaging using synchrotron radiation, necessitating higher temporal resolutions. Consequently, this places stringent demands on X-ray detector technology, especially when X-ray energy exceeds \SI{20}{keV}. 
The selection of a suitable scintillator material is crucial for achieving optimal timing resolution, yet it poses a significant challenge in dynamic X-ray imaging.
This study delves into the optimization of scintillator properties and their impact on spatial resolution and light output, elucidating the performance of Ce-doped Gd$_3$Ga$_3$Al$_2$O$_{12}$ (GAGG:Ce) scintillators for X-ray imaging applications. 
We developed a micro X-ray imaging detector using a \SI{100}{\micro \meter}-thick GAGG:Ce scintillator plate and conducted X-ray imaging tests at the Aichi SR facility. 
The results demonstrated that the resolution, quantified as the chart slit width at a contrast transfer function (CTF) value of 10\%, reached \add{2 $\sim$ \SI{3}{\micro \meter}} with a 4x lens, \add{\SI{0.52}{\micro \meter} $\pm$ \SI{0.03}{\micro m}} with a 20x lens, and \add{\SI{0.42}{\micro \meter} $\pm$ \SI{0.01}{\micro m}} with a 40x lens. 
\add{
Although the results of this study did not achieve a spatial resolution nearing the effective pixel size of the 40x lens, the text also elucidates the underlying reasons for this limitation.
}
\erase{
Notably, this resolution remained nearly identical for both GAGG:Ce and commercially available Lu$_3$Al$_5$O$_{12}$:Ce (LuAG:Ce) scintillator plates with a consistent thickness of 100~$\mu$m. 
}
Furthermore, we compared the X-ray sensitivity of our GAGG:Ce scintillator plate with that of a commercial LuAG:Ce scintillator, revealing an approximately 1.5-fold increase in light output. 
As a demonstration, transmission images of dried small fish were captured using the GAGG:Ce scintillator plate and the developed X-ray imaging system. 
These findings highlight the potential of the X-ray imaging detector devised in this study for future generations of X-ray imaging applications. 
}


\begin{document}
\maketitle
\flushbottom

\section{Introduction}
\label{sec:intro}
Scintillator-based X-ray imaging detectors, recognized as a mainstream choice in today’s marketplace, assume a pivotal role in scientific inquiry and practical applications encompassing medical imaging, non-destructive quality inspection of food and devices, as well as security monitoring \cite{Gruner2002-du,Martin2006-wz,Tous2013-oh,Yoshino2023-pk,Hatsui2015-yk}.
Recent years have witnessed an emergence of studies focusing on \add{4D (time-resolved 3D)} \erase{4D} X-ray imaging at the millisecond scale utilizing synchrotron radiation \cite{Yashiro2017-dc,Yashiro2022-ix,Liang2023-kr,Kawanishi2023-qc}. 
\add{
4D X-ray imaging, offering millisecond time resolution and micrometer spatial resolution, has a broad spectrum of applications. These range from academic research, such as the observation of biological processes in insects and the analysis of material degradation, to industrial uses, including the development of intelligent materials and the exploration of dynamic biomimetic applications.
}
As the demand for higher resolution at shorter time intervals in 4D X-ray imaging experiments increases, the requirements for X-ray detector technology are becoming increasingly stringent.
\erase{As the demand for higher resolution and shorter timings in 4D X-ray imaging experiments intensifies, the requirements on X-ray detector technology are becoming increasingly stringent.}
\erase{
Consequently, constraints are imposed on achieving shorter timings and higher spatial resolutions in such experiments.
}
Notably, \erase{time-resolved} 4D X-ray imaging becomes more complex when the X-ray energy exceeds \SI{20}{keV}. 
Therefore, the selection of an appropriate scintillator is crucial in determining the achievable temporal resolution and serves as a critical bottleneck in the advancement of dynamic 4D X-ray imaging.

\add{
The light emitted by the scintillator screen upon interaction with X-rays radiates omnidirectionally within the scintillation screen, starting from the point of interaction. 
Scintillation light propagating laterally can introduce blurring in imaging systems that utilize optical lens coupling. 
Consequently, minimizing lateral light scattering within the scintillation screen is fundamentally important for enhancing the spatial resolution of X-ray images.
Previous studies\cite{Xie2016-th,Alikunju2023-zi} have explored the relationship between scintillator screen thickness and spatial resolution, revealing that the spatial resolution of X-ray images tends to degrade as the screen thickness increases.
}
\erase
{In principle, minimizing lateral light scattering within the scintillation screen is desirable to enhance the spatial resolution of X-ray imaging.}
Furthermore, a study has been documented wherein the thickness of the scintillator was reduced to \SI{5} {\micro \meter}, achieving a spatial resolution of \SI{200} {\nano \meter} for X-ray imaging \cite{Kameshima2019-zq}.
Conversely, to enhance scintillation brightness, which directly influences the sensitivity of an X-ray imaging detector, the scintillation screen must exhibit adequate thickness. Consequently, there exists a trade-off between sensitivity (scintillation brightness) and spatial resolution in X-ray imaging, underscoring the significance of ascertaining the scintillation screen thickness that optimizes this trade-off in X-ray imaging applications. 
For highly transparent single crystal scintillators, spatial resolution should remain constant for a given thickness, irrespective of the scintillator material. Subsequently, the sensitivity of X-ray imaging is profoundly contingent upon the scintillator material, and can be expressed as follows:

\begin{equation}
    \mathrm{Sensitivity}\propto\mathrm{LY} \times E_{dep},
\end{equation}
   \begin{equation}
       E_{dep} = \int S(E)P(E)\, dE,
   \end{equation}
\begin{equation}
    P(E) = 1-\exp(-\mu_m (E,Z) \rho d).
\end{equation}

Where $\mathrm{LY}$ denotes the light yield of the scintillator, $E_{dep}$ represents the deposited energy by the scintillator, $S(E)$ denotes the energy distribution function of X-rays emitted from an X-ray tube, $P(E)$  represents the interaction probability density function between scintillator and X-rays, $\mu_m(E,Z)$ stands for the mass attenuation coefficient (unit: \si{cm^2/g}) of the scintillator, $Z$ denotes the effective atomic number of the scintillator, $\rho$ represents the density (unit: \si{g/cm^3}) of the scintillator, and $d$ signifies the thickness of the scintillator, respectively. 
The sensitivity of X-ray imaging is intricately linked to various factors, including the scintillator's light yield, density, effective atomic number, and thickness.

Ce-doped Lu$_3$Al$_5$O$_{12}$ (LuAG:Ce)  has been the primary candidate material for X-ray imaging scintillators because of its light yield of \SI{12500} {ph/MeV}, fast decay time of \SI{60} {ns}, high density of \SI{6.67} {g/cm^3}, effective atomic number of 58.9, and emission wavelength of \SI{540} {nm}, which is very compatible with the quantum efficiency of CMOS sensors\cite{Nikl2013-ne,Chewpraditkul2009-px,Mares2004-tc}. 
The theoretical light yield estimated by Drenbos et al. \cite{Dorenbos2010-uk} amounts to \SI{60000}{ph/MeV}; however, only a fraction, less than one-fourth, of this value has been empirically reported.
This is due to the presence of a very long scintillation decay component on the order of milliseconds in the LuAG:Ce scintillation process. This long decay component delays the radiative recombination of free charge carriers through the luminescent center, Ce$^{3+}$, as these charge carriers are retrapped by shallow electron traps. This slow scintillation decay time on the order of milliseconds is also a problem in 4D X-ray CT where millisecond time resolution is required.
Ce-doped Gd$_3$Ga$_3$Al$_2$O$_{12}$ (GAGG:Ce) epitomizes a garnet-type structural scintillator, boasting a commendable light yield of \SI{50000}{ph/MeV}, fast decay time of \SI{90}{ns}, high density of \SI{6.63}{g/cm^3}, high effective atomic number of 55, and emission wavelength of \SI{550}{nm} \cite{Kamada2011-mh,Kamada2016-kg,Rawat2018-qg}.
In this study, GAGG:Ce single crystals with a diameter of 2-inch were grown using the Czochralski method. 
The resultant GAGG:Ce single crystal was utilized to fabricate a scintillator plate for X-ray imaging, with a thickness of \SI{100}{\micro \meter}. 
An X-ray imaging detector was then constructed using this plate, and its X-ray imaging performance was compared to that of a commercially available LuAG:Ce scintillator plate. 

\section{Materials and Methods}
\label{sec:mate}
The 2-inch diameter GAGG:Ce single crystals were grown using the Czochralski method with a mixture of Gd$_2$O$_3$, Ga$_2$O$_3$, Al$_2$O$_3$, and CeO$_2$ in a (Gd$_{0.995}$, Ce$_{0.005}$)$_3$Ga$_3$Al$_2$O$_{12}$ composition ratio.
The GAGG:Ce scintillator plates used in this study were prepared by cutting and polishing a \SI{5}{mm} $\times$ \SI{5}{mm} $\times$ \SI{100}{\micro \meter}-thick from a 2-inch diameter GAGG:Ce single crystal.
To evaluate the imaging performance of the GAGG:Ce scintillator plate, we constructed an X-ray imaging system equipped with magnifying optics. 
The X-ray imaging system employed in this study comprises three primary components: 
\begin{enumerate}
    \item The scintillation segment, encompassing a scintillation screen tasked with the conversion of X-rays into visible light. Both GAGG:Ce synthesized in this study and commercially available LuAG:Ce (Hamamatsu Photonics, Hamamatsu, Japan) were employed. A standardized thickness of \SI{100}{\micro \meter} was adopted.
    \item infinity-corrected optics, designed to magnify the scintillation light and project it onto a CMOS sensor. Objective lenses from the 4x, 10x, and 20x Plan Apochromato series (Nikon Solutions Co., Ltd., Tokyo, Japan) were employed.
    \item the sensor segment, where the magnified scintillation light is captured and sampled by the cooled CMOS sensor. A CS67-M monochrome cooled CMOS sensor (BITRAN CORPORATION, Saitama, Japan) with a pixel size of \SI{9}{\micro \meter} $\times$ \SI{9}{\micro \meter} and \SI{1604}{pix} $\times$ \SI{1100}{pix} effective pixels was employed. The image field of view sizes with 4x, 10x, and 20x objective lenses are \SI{3610}{\micro m} $\times$ \SI{2740}{\micro m}, \SI{1440}{\micro m} $\times$ \SI{990}{\micro m}, and \SI{720}{\micro m} $\times$ \SI{495}{\micro m}, respectively.
\end{enumerate}

\begin{figure}[htbp]
    \centering
    \includegraphics[width=1\linewidth]{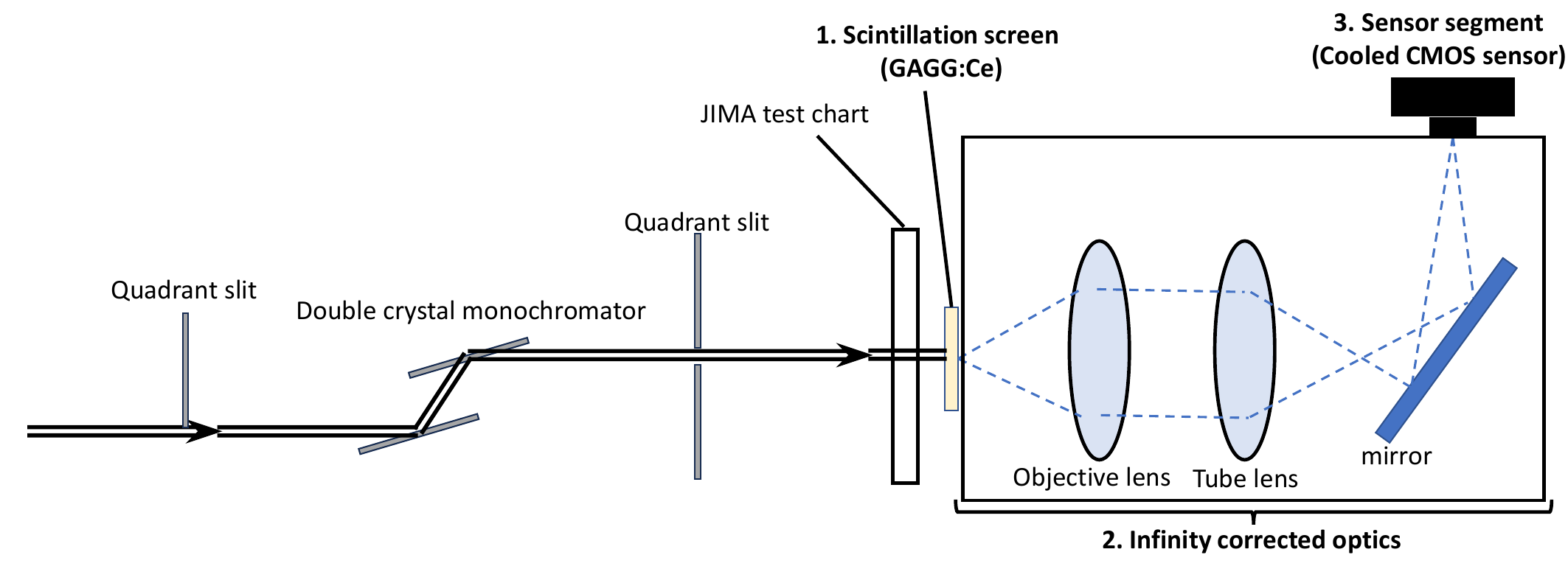}
    \caption{Schematic of the X-ray imaging experimental setup and optical configuration of the detector.}
    \label{fig:setup}
\end{figure}

\add{
A schematic representation of the X-ray imaging test is presented in Fig. \ref{fig:setup}.
}
Beamline BL8S2 at the Aichi synchrotron radiation center (AichiSR \cite{web-aichiSR}) was used as the X-ray source for the X-ray imaging test. 
Continuous X-rays with photon energies ranging from \SIrange{7}{24}{keV} were irradiated under a nitrogen atmosphere at a photon velocity density of \SI{6.2e8}{Photons\per\second\per\square\mm} @ \SI{9.8}{keV}.

Spatial resolution evaluation was performed using a micro resolution chart for X-Ray (RT RC-04, Japan Inspection Instruments Manufacturers' Association(JIMA), Tokyo, Japan). 
Three transmitted images of the micro resolution chart were captured within the identical field of view and subsequently averaged; each frame was exposed for 10 ms with the x4 lens, 100 ms with the 20x lens, and 400 ms with the 40x lens. Subsequently, flat field correction was applied to the acquired images, generating a corrected image by dividing the transmitted chart image by a blank image captured under comparable imaging conditions.
The Contrast Transfer Function (CTF) serves as a pivotal parameter in evaluating the spatial resolution performance of an imaging system; 
it offers a quantitative and standardized characterization of said system, illuminating how object surface contrast manifests in the image plane. 
A transmission line profile was derived from the X-ray images depicting these linear patterns. To derive the CTF, the contrast $c$ was initially calculated, defined as the ratio of intensities between open and covered areas:
\begin{equation}\label{eq:contrast}
    c = \frac{I_{max} - I_{min}}{I_{max} + I_{min}}
\end{equation}
where $I_{max}$ and $I_{min}$ denote the mean intensities of maxima and minima, respectively, for each spatial frequency. 
As spatial frequency escalates, contrast diminishes, enabling determination of the smallest resolvable structure size. 
However, the contrast calculated in Equation \ref{eq:contrast} necessitates normalization by the maximum contrast $c_{max}$, representing the contrast between fully open and fully covered areas. 
This yields the CTF value for each spatial frequency:
\begin{equation}
    CTF(f) = \frac{c(f)}{c_{max}}
\end{equation}
Spatial resolution of the imaging system $r_{CTF}$ is conventionally defined as the inverse of the spatial frequency at the 10\%-level of the CTF:
\begin{equation}
    r_{CTF} = f^{-1}|_{CTF=10\%}
\end{equation}
While CTFs lack a quantitative expression of spatial resolution, their transmission profiles offer an approximation of a system's imaging capabilities. 
\add{
The primary potential sources of error in this study include temporal fluctuations in synchrotron X-ray intensity and geometric variations in pixel values, attributable to the finite size of the camera pixels and the amount of light entering each pixel. 
To mitigate these error factors as much as possible, three X-ray transmission images were acquired for each measurement. Additionally, 7 regions of interest were defined for each slit in the images, and the CTF was calculated to determine the mean and measurement error.
}

In X-ray imaging, the light output of scintillation light is determined by parameters such as the amount of scintillation light yield per unit X-ray energy (photons/keV), the interaction probability between the incident X-rays and the scintillator, and the transmittance of the scintillator.
\add{
In this study, we assume that the average pixel value (in ADU) of a blank region in an X-ray transmission image is proportional to the light output of the scintillation screen. We then compare the average pixel values when different scintillators are replaced within the same imaging system to evaluate the scintillation screen's light output.
Since CCD cameras register positive pixel values even in the absence of X-ray exposure due to thermal noise and other factors, we measured X-ray transmission images at varying exposure times and plotted the relationship between exposure time and average pixel value. The resulting data were fitted with a linear function, and the slopes were compared as an index of the scintillation screen’s light output. 
To ensure accuracy, three X-ray transmission images were taken for each exposure time, six regions of interest were selected from each image, and the average pixel value was calculated based on the data from these regions of interest.
The exposure time was determined according to the intensity of the synchrotron radiation X-rays, the dynamic range of the CCD, and the minimum exposure time of 1 ms. 
The CCD employed in this study is capable of recording pixel values with a 12-bit resolution. The exposure time was adjusted to ensure that the pixel value did not exceed the maximum threshold of 4096.
}
\erase
{In this study, the average pixel value (unit: ADU) of a CCD camera in blank areas of X-ray transmission images was compared to the scintillation light output.}

\section{Results and discussion}
\label{sec:result}

\begin{figure}[htbp]
\centering
\begin{minipage}[t]{0.5\linewidth}
    \includegraphics[keepaspectratio, width=1\linewidth]{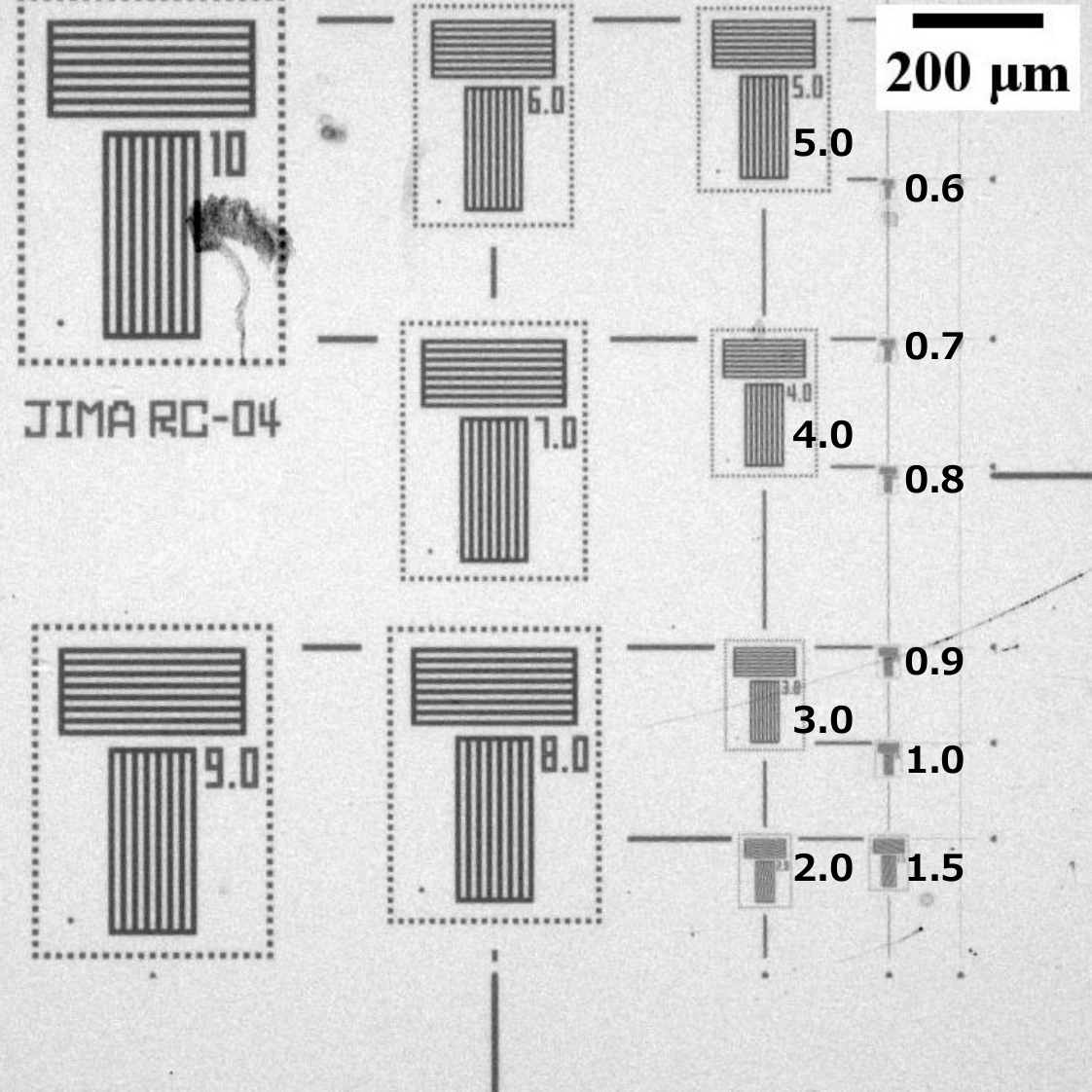}
{\captionsetup{position=bottom,justification=centering} \subcaption{}}\label{subfig:X-ray-image_4x}
\end{minipage}
\vskip3mm
\hspace{-1cm}

\begin{minipage}[t]{0.5\linewidth}
    \includegraphics[keepaspectratio, width=1\linewidth]{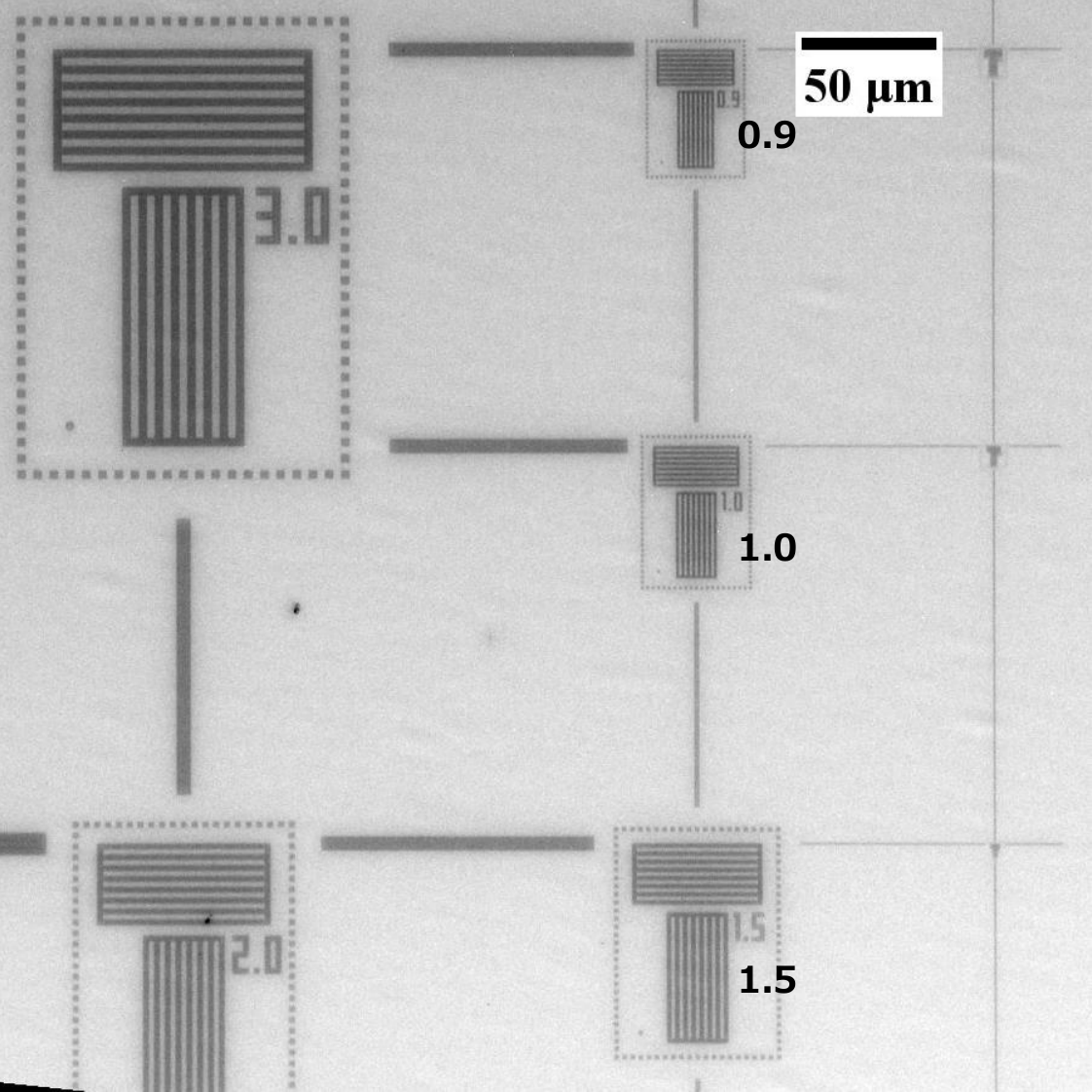}
{\captionsetup{position=bottom,justification=centering} \subcaption{}}\label{subfig:X-ray-image_20x}
\end{minipage}
\vskip3mm
\hspace{-1cm}

\begin{minipage}[t]{0.5\linewidth}
    \includegraphics[keepaspectratio, width=1\linewidth]{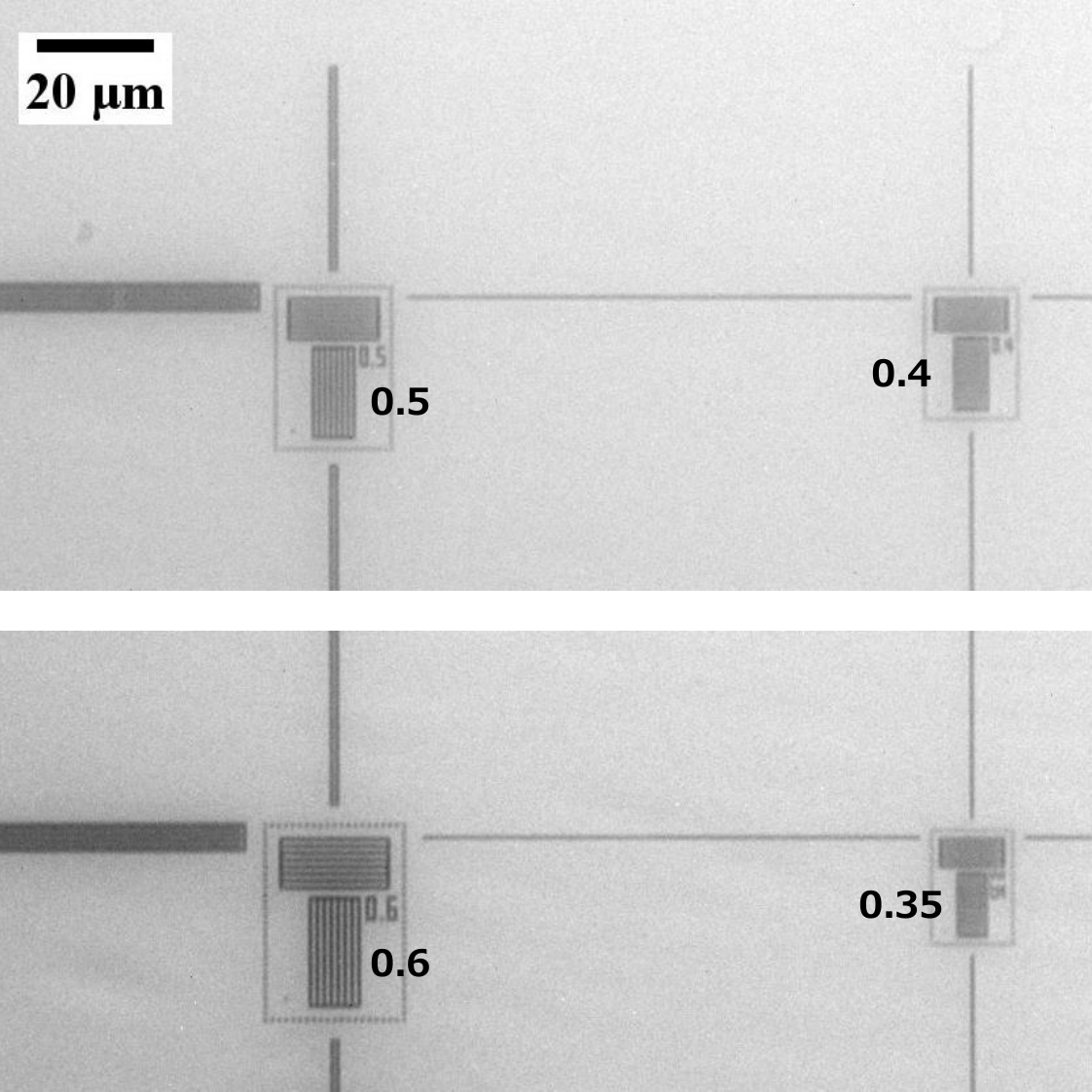}
{\captionsetup{position=bottom,justification=centering} \subcaption{}}\label{subfig:X-ray-image_40x}
\end{minipage}
    \caption{X-ray transmission image of JIMA chart taken with \textbf{(a)} 4x, \textbf{(b)} 20x, and \textbf{(c)} 40x objective lenses.}
    \label{fig:X-ray-image}
\end{figure}

Fig. \ref{fig:X-ray-image} \textbf{(a)}, \textbf{(b)}, and \textbf{(c)} display flat-field corrected X-ray transmission images of the JIMA chart captured using 4x, 20x, and 40x objective lenses, respectively.
The numerical annotations adjacent to the JIMA microchart in the images denote the line widths of the chart.
The chart images reveal that the line width can be resolved to approximately \SI{3.0}{\micro m} in the X-ray image acquired with the 4x lens and to about \SI{0.5}{\micro m} in the X-ray image obtained with the 40x lens.

\begin{figure}[htbp]
    \centering
    \includegraphics[width=1\linewidth]{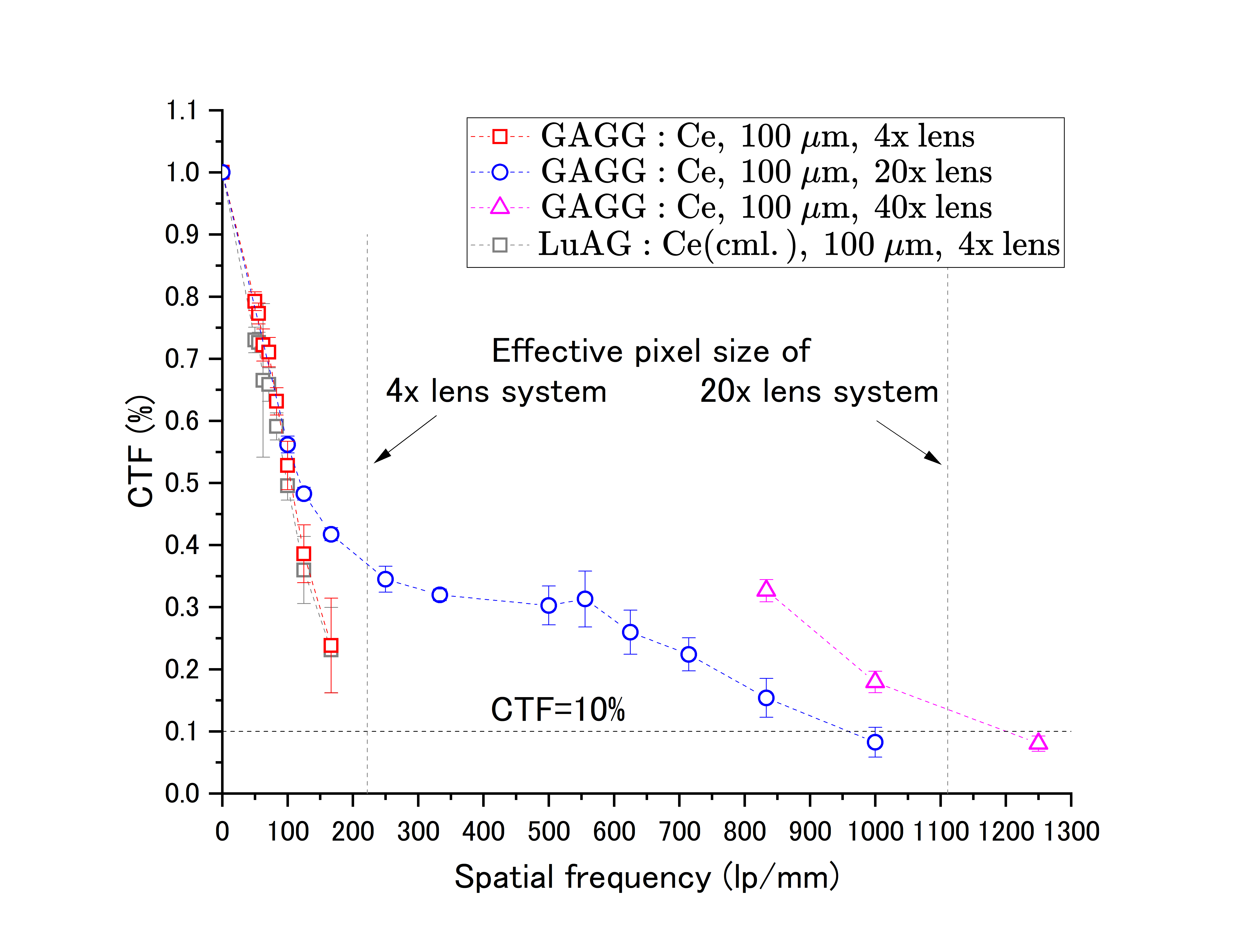}
    \caption{Calculated CTF value as a function of spatial frequency (lp/mm) for 4x, 20x, and 40x objective lenses.}
    \label{fig:ctf}
\end{figure}

The results of CTF calculations on X-ray transmission images taken with 4x, 20x, and 40x objective lenses are shown in Fig. \ref{fig:ctf}.
The horizontal black dotted line depicting a CTF value of 10\%-level is also delineated in Fig. \ref{fig:ctf}.
\add{
In the calculation of the CTF value for X-ray images obtained using a 4x lens, it was feasible to compute the CTF value for slits as narrow as \SI{3}{\micro m}, as the number of slits corresponded to the number of line profile peaks.
However, for slits narrower than \SI{2}{\micro m}, the number of line profile peaks fell below the number of slits, making it impossible to calculate the CTF value.
When using a 4x objective lens, the effective pixel size of the detector is \SI{2.25}{\micro m}, suggesting that slits narrower than \SI{2}{\micro m} cannot be resolved in the line profile.
}
\erase{
Fig. \ref{fig:ctf} illustrates that the spatial frequency at which the CTF value attains 10\% for the X-ray image captured with the 4x objective is approximately 228~lp/mm, corresponding to a line width of 2.2~$\mu$m. 
Considering the effective pixel size of the detector when employing the 4x objective is 2.25~$\mu$m, it indicates that the spatial resolution is nearly equivalent to the effective pixel size of the 4x lens system.
}
In Fig. \ref{fig:ctf}, the X-ray image captured with a 20x objective lens exhibits a CTF of 10\% at a spatial frequency of 
\add{\SI{950}{lp/mm} $\pm$ \SI{60}{lp/mm}}
, corresponding to a line width of 
\add{\SI{0.52}{\micro \meter} $\pm$ \SI{0.03}{\micro \meter}}.
The X-ray image taken with a 40x objective lens shows a spatial frequency of \add{\SI{1200}{lp/mm} $\pm$ \SI{30}{lp/mm}}, corresponding to a line width of \add{\SI{0.42}{\micro \meter} $\pm$ \SI{0.01}{\micro \meter}} at a CTF of 10\%. 
The vertical dashed lines in Fig. \ref{fig:ctf} indicate the spatial frequencies corresponding to the effective pixel sizes of the detector (\SI{2.25}{\micro \meter} and \SI{0.45}{\micro \meter}) when using the 4x and 20x objective lenses, respectively. 
\erase{
The X-ray image captured with the 4x objective provides a spatial resolution comparable to the effective pixel size, while the image captured with the 20x objective results in a spatial resolution worse than the effective pixel size. 
}
The resolution of X-ray images taken with 4x and 20x objectives approached the effective pixel size. 
\add{
Conversely, X-ray imaging using a 40x lens did not achieve a resolution approaching the effective pixel size of \SI{0.225}{\micro m}, which corresponds to \SI{2222}{lp/mm}.
}
\erase{
Conversely, with the 40x objective, the effective pixel size was 0.225~$\mu$m, and imaging with the 40x lens did not achieve resolution at the effective pixel size. 
}

The precise reason for the spatial resolution of X-ray images taken with the 40x lens not matching the effective pixel size remains undetermined. 
\add{
However, previous studies have reported several factors contributing to the degradation of spatial resolution.
The first factor is the fundamental limit of optical resolution. The primary constraint on optical device resolution is the wavelength of the light used in imaging experiments.
When light of wavelength $\lambda$ propagates through a medium with a refractive index $n$ and converges to a focal point with a half-angle $\theta$, the minimum resolvable distance $d$ is given by the following equation:
\begin{equation}\label{eq:resolution limit}
    d = \frac{\lambda}{2n sin\theta} = \frac{\lambda}{2NA}
\end{equation}
, where NA is the numerical aperture.
In this study, the NA of the 40x objective lens is 0.95, and the emission wavelength of GAGG:Ce is \SI{540}{nm}, yielding a minimum resolvable distance of $d = \SI{284}{nm}$.
This does not account for the resolution degradation observed in the experimental results.
}

\add{
The second potential factor is related to X-ray energy.
In a report by Kameshima et al. \cite{Kameshima2019-zq}, the resolution was shown to deteriorate at higher X-ray energies, with an investigation covering energies from \SI{7.4}{keV} to \SI{18}{keV}.
In this study, white X-rays with a maximum energy of \SI{24}{keV} were utilized, which likely contributed to the observed decline in resolution.
}

\add{
The third possible cause pertains to the scintillator thickness. 
Previous studies \cite{Xie2016-th,Alikunju2023-zi} have explored the relationship between scintillator thickness and resolution.
In this study, a scintillator screen with a thickness of \SI{100}{\micro m} was employed, which is significantly thicker than the \SI{5}{\micro m} scintillator used in the Kameshima et al. \cite{Kameshima2019-zq} that reported a resolution of \SI{200}{nm}.
It is probable that the increased thickness of the scintillator contributed to the resolution deterioration in the present study.
A thinner scintillator screen could theoretically provide higher spatial resolution, but there is a trade-off in terms of reduced X-ray detection efficiency.
Therefore, the optimal thickness of the scintillator screen for X-ray imaging must be determined by balancing factors such as X-ray energy and the desired resolution.
}

\erase
{However, Kameshima et al. \cite{Kameshima2019-zq} have identified a relationship between X-ray energy and resolution, with higher energies leading to resolution degradation.
In this study, continuous X-rays with a maximum energy exceeding 24~keV were used, which may have contributed to the observed resolution degradation.
}

\begin{figure}[htbp]
    \centering
    \includegraphics[width=1\linewidth]{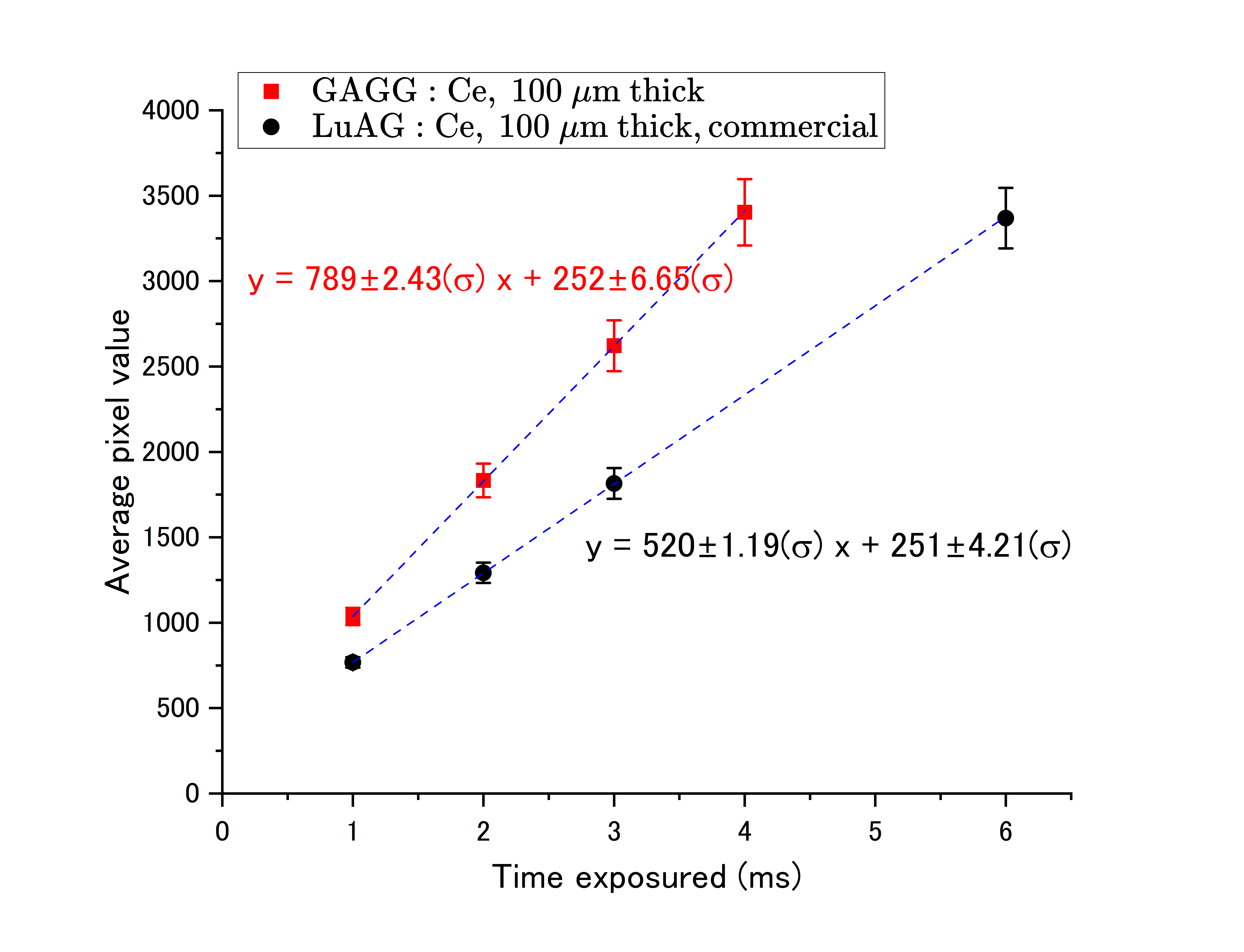}
    \caption{Average pixel values in X-ray images of a blank region for \textbf{(red square)} \SI{100}{\micro \meter} thick GAGG:Ce and \textbf{(black circle)} \SI{100}{\micro \meter} thick LuAG:Ce.}
    \label{fig:pixelvalue-vs-time}
\end{figure}

\begin{figure}[htbp]
    \centering
    \includegraphics[width=1\linewidth]{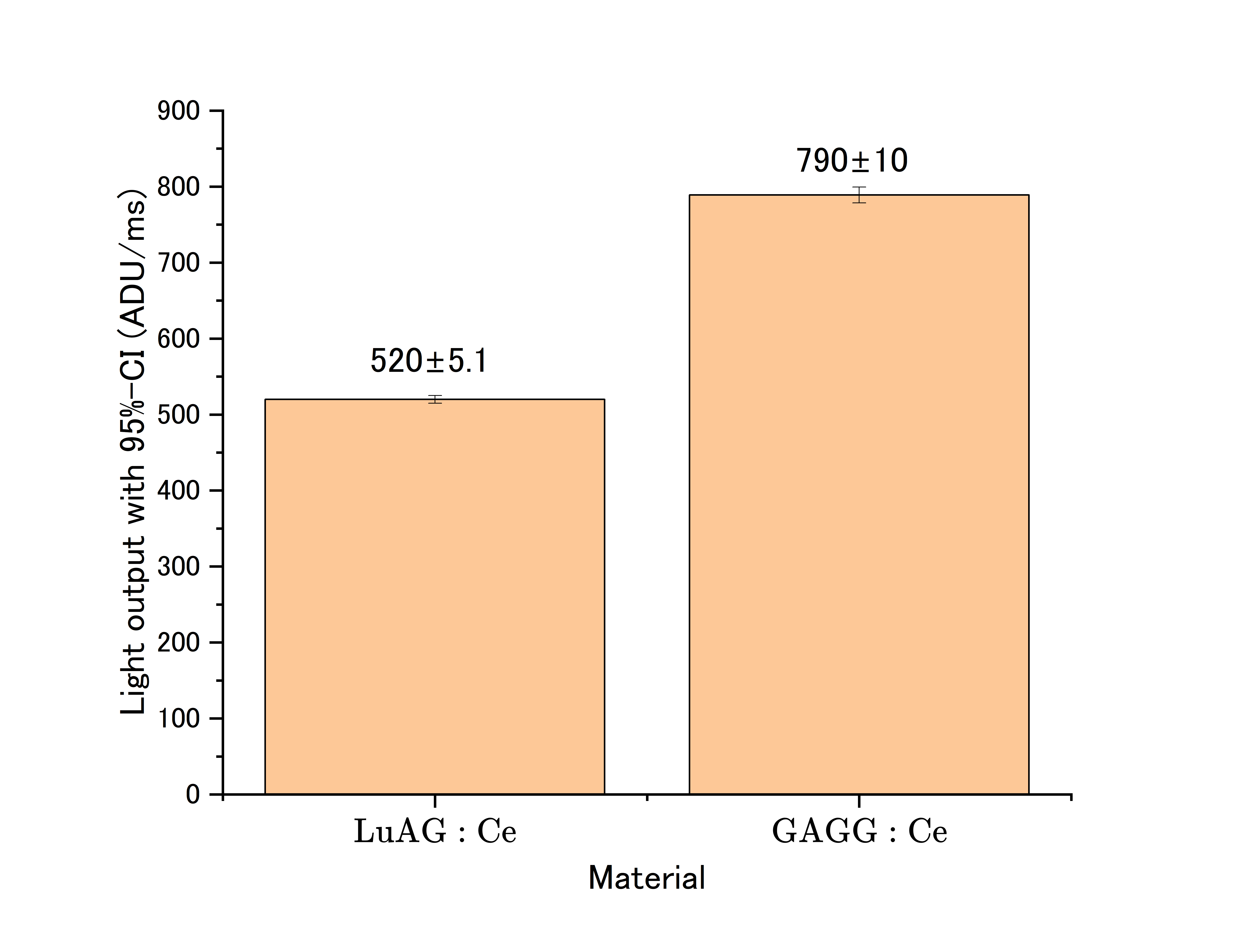}
    \caption{Comparison of light output values for \SI{100}{\micro \meter} thick LuAG:Ce (commercial) and \SI{100}{\micro \meter} thick GAGG:Ce scintillators.}
    \label{fig:lightoutput}
\end{figure}

The relationship between the average pixel value and X-ray exposure time is shown in Fig. \ref{fig:pixelvalue-vs-time}. 
It can be observed that, at the same exposure time, GAGG:Ce exhibits a higher average pixel value compared to LuAG:Ce. 
The scintillation light output in X-ray imaging detectors can be expressed as the slope in the relationship between average pixel value and X-ray exposure time. 
Fig. \ref{fig:lightoutput} illustrates a comparison of scintillation light output values between GAGG:Ce and LuAG:Ce. As a result, it was found that the scintillation light output of GAGG:Ce is 1.5 times higher than that of LuAG:Ce.
The observed 1.5-fold difference in light output is lower compared to the difference calculated based on light yield (\SI{50000}{ph/MeV} and \SI{12500}{ph/MeV}), density (\SI{6.63}{g/cm^3} and \SI{6.67}{g/cm^3}), and effective atomic number (55 and 58.9) discrepancies between GAGG:Ce and LuAG:Ce.
This discrepancy might stem from the remarkably slow luminescence kinetics of LuAG:Ce, which decays on a millisecond scale. A previous study by Nikl et al. \cite{Nikl2013-ne} proposed that the retrapping of migrating electrons in shallow traps associated with cation antisite defects during the transport stage of the scintillation mechanism constitutes the primary cause of the very slow scintillation observed in Ce-doped aluminum garnet scintillators.
While the light yield of LuAG:Ce measured within the \SI{1}{\micro \second} time window is as low as \SI{12500}{ph/MeV}, the integrated radioluminescence spectrum is notably high, reaching 700\% of that of BGO.
The light output quantified in this study was assessed using pixel values extracted from blank images captured with CMOS sensors, employing exposure times ranging from \SIrange{1}{4}{\milli \second}, thereby suggesting that the millisecond-order slow scintillation decay of LuAG:Ce is indeed reflected in the heightened light output.

\add{
Notwithstanding these considerations, the \SI{100}{\micro m} thick GAGG:Ce scintillator screen developed in this study achieved a light output 1.5 times greater when compared to the commercially available LuAG:Ce screen and demonstrated a good spatial resolution of \SI{0.42}{\micro m}. In the advancement of next-generation X-ray CT systems, the enhanced spatial resolution facilitates the detection of finer objects, while the increased light output allows for expedited CT imaging by producing brighter images even during shorter exposure times. In this context, the findings of this research are poised to significantly contribute to the evolution of next-generation X-ray CT systems. Notably, the low light output of scintillator screens has been a critical impediment to the realization of 4D X-ray CT; thus, the results of this study will assist in addressing this challenge.
}
\erase{
Notwithstanding these considerations, it is noteworthy that the light output of GAGG:Ce persists at a level 1.5 times greater than that of LuAG:Ce.
}

\begin{figure}[htbp]
    \centering
    \includegraphics[width=1\linewidth]{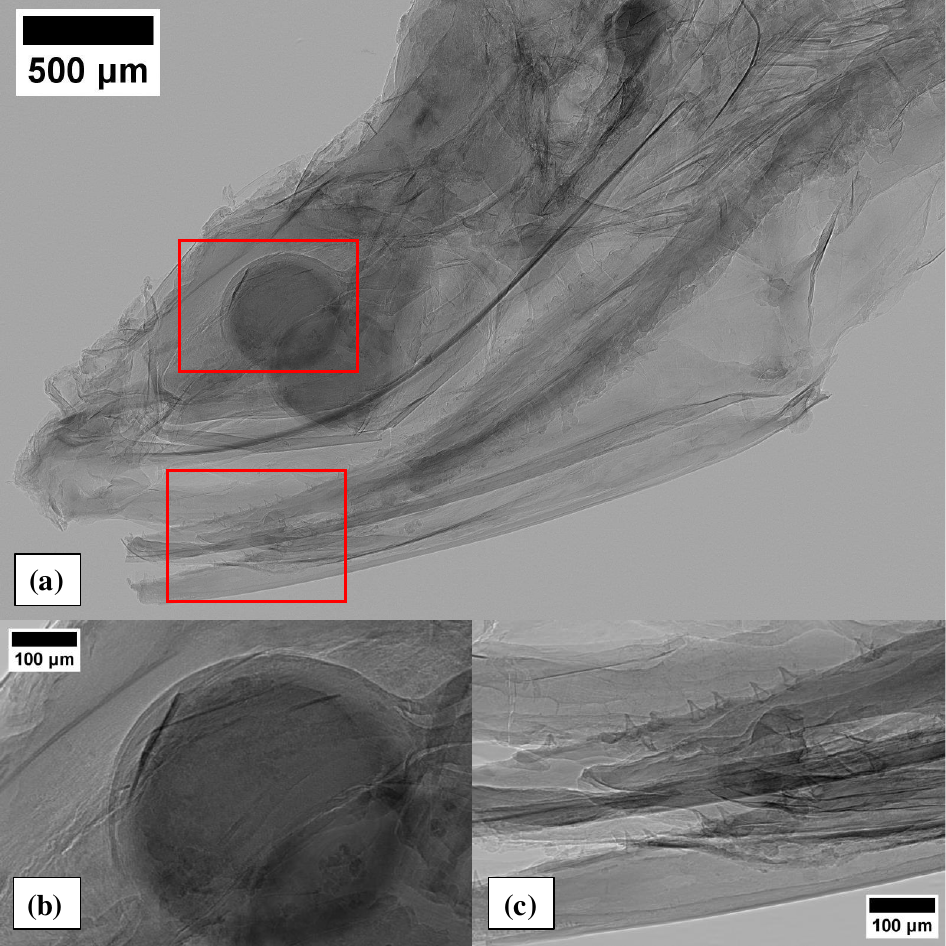}
    \caption{\textbf{(a)} X-ray image of desiccated small fish captured using a 4x objective lens.  \textbf{(b), (c)} Magnified X-ray images of the red square in Fig. 6 \textbf{(a)} obtained with a 20x lens.}
    \label{fig:sample}
\end{figure}

Figure \ref{fig:sample} \textbf{(a)} presents an X-ray transmission image of a desiccated fish captured with a 4x objective lens. Figures \ref{fig:sample} \textbf{(b)} and \textbf{(c)} display magnified X-ray transmission images of the red square region in Figure \ref{fig:sample} \textbf{(a)}, obtained using a 20x lens.
These results demonstrate that X-ray imaging accurately reveals the internal anatomical structure of the dried small fish. Figures \ref{fig:sample} \textbf{(b)} distinctly shows the ocular lens within the dried small fish, while Figures \ref{fig:sample} \textbf{(c)}  clearly depicts the mouth and dentition of the dried small fish.
Based on these findings, the GAGG:Ce based X-ray imaging detector developed in this study is anticipated to be effectively utilized as a synchrotron radiation X-ray imaging detector.

\section{Conclusion}
\label{sec:conc}
The selection of scintillator material stands as a pivotal determinant in investigations necessitating exceedingly sub-millisecond timing resolution, such as 4D X-ray imaging employing synchrotron radiation.
In this investigation, we constructed an X-ray imaging detector utilizing a \SI{100}{\micro \meter} thick GAGG:Ce scintillator, subsequently assessing its spatial resolution and light output. 
The attained spatial resolution, derived from the calculated CTF values, reached \add{2 $\sim$ \SI{3}{\micro m}} with a 4x objective lens, \SI{0.52}{\micro m} with a 20x objective lens, and  \SI{0.42}{\micro m} with a 40x objective lens.
Concerning light output, we juxtaposed the GAGG:Ce scintillator against a commercially available \SI{100}{\micro \meter}-thick LuAG:Ce scintillator, unveiling a 1.5-fold increase in light output with the GAGG:Ce.
With 4x and 20x objectives, the resolution of X-ray imaging can achieve values close to the effective pixel size. 
However, even with the 40x lens, it was not possible to resolve slits finer than \SI{0.4}{\micro m}. 
\add{
This limitation may arise from the high energy of the continuous X-rays employed in this study (\SI{24}{keV}) and the substantial thickness of the scintillator (\SI{100}{\micro m}). 
Future investigations should concentrate on enhancing this experimental constraint on spatial resolution without compromising the scintillator's light output. 
To achieve this, it is essential to calculate the X-ray attenuation length at the employed energy and determine the minimum thickness that prevents X-ray penetration into the scintillator screen.
Alternatively, utilizing scintillator screens with superior light output will also be crucial for achieving both high temporal and spatial resolution. 
Recent studies have indicated that garnet-type scintillators co-doped with Ce and Tb demonstrate greater light output than GAGG:Ce \cite{Omuro2024-dx,Omuro2024-ce}. 
Employing these scintillator screens could significantly address the aforementioned challenges.
}
\erase{
This limitation may be attributed to the high energy of the continuous X-rays used in this study, which was 24~keV.
}
The findings of this study hint at the potential utility of the developed GAGG:Ce-based X-ray imaging detector for practical application in dynamic X-ray imaging employing synchrotron radiation, provided the aforementioned issue are addressed.


\acknowledgments
This work was conducted at the BL8S2 of Aichi Synchrotron Radiation Center, Aichi Science \& Technology Foundation, Aichi, Japan (Proposal No.202305163 and No.202306133) with the financial support of Synchrotron Radiation Research Center, Nagoya University.
This work was supported by the JSPS KAKENHI  [grant numbers JP21H03834, JP21KK0082, JP18H01222, JP21K07699, JP19H00672, JP22H03019].
This work was based on results obtained from a project,
JPNP20004, subsidized by the New Energy and Industrial Technology Development
Organization (NEDO).
This work was supported by JST ERATO Grant Number JPMJER2102, Japan.


\bibliographystyle{JHEP}
\bibliography{biblio.bib}

\end{document}